\documentclass[12pt]{article}
\usepackage{amsfonts}
\def\be{\begin{equation}}
\def\ee{\end{equation}}
\def\la{\label}
\def\ci{\cite}
\def\bi{\bibitem}
\def\<{\langle}
\def\>{\rangle}
\def\vf{\varphi}
\def\b{\begin{equation}}
\def\e{\end{equation}}
\def\be{\begin{eqnarray}}
\def\ee{\end{eqnarray}}
\def\R{\mathbb R}
\def\C{\mathbb C}
\def\F{{\bf \Phi}}
\def\H{{\cal H}}

\begin{document} 

\begin{center}

{\LARGE{\bf   Gamow Functionals on Operator Algebras.}}   

\vskip1cm

{\large {\bf M. Castagnino$^\dagger$, M. Gadella$^{\dagger\dagger}$, R. Id
Bet\'an$^\dagger$, R. Laura$^\dagger$.  }} 

\end{center}

\vskip1cm
$^\dagger$Facultad de Ciencias Exactas, Ingenier\'{\i}a y Agrimensura. Av.
Pellegrini 250, (2000) Rosario, Argentina.  

$^{\dagger\dagger}$Departamento de F\'{\i}sica Te\'orica. Facultad de Ciencias.
c./ Real de Burgos, s.n. 47011 Valladolid, Spain.

\begin{abstract}

We obtain the precise form of two Gamow functionals, representing the
exponentially decaying part of a quantum resonance and its mirror image that
grows exponentially, as a linear, positive and continuous functional on an
algebra containing observables. These functionals do not admit normalization
and, with an appropiate choice of the algebra, are time reversal of each other.

\end{abstract}

\centerline{\bf\today}

\section{Introduction.}

The goal of the present paper is to give a precise definition of the Gamow
functional on a formalism that has been used previously to discuss a variety of
topics such as resonance behaviour, decoherence, generalized states with
diagonal singularity, etc \ci{LC1,LC2,LC3,LC4,VH}. This formalism has been
inspired in previous work by Prigogine and collaborators \ci{A1,PP,8}. 

Gamow vectors \ci{G} are generalized eigenvectors of the total Hamiltonian, 
in a resonant scattering process, with complex eigenvalues given by the
simple poles of the analytic continuation of the $S$-matrix \ci{B} or the
reduced resolvent
\ci{HM,AP,A2,AM}. As the Hamiltonian is a self-adjoint operator, its
eigenvectors with complex eigenvalues cannot live in a Hilbert space but on
certain extensions of the Hilbert spaces: the rigged Hilbert Spaces (RHS)
\ci{B,BG,AP,A2}. Gamow vectors represent the exponentially decaying part of a
resonance (for a discussion on the decay in quantum mechanics see \ci{FGR}). The
question arises of whether a Gamow vector represents a truly quantum state
i.e., an element of the physical reality. 

In conventional quantum mechanics in Hilbert space, let $|\vf\>$ be a pure
state. Its corresponding density operator is given by $\rho=|\vf\>\<\vf|$. The
operator $\rho$ represents the state $|\vf\>$ in the Liouville space and,
therefore, this is the object that should represent the state in quantum
statistical mechanics. Thus, if we accept that the Gamow vector represents a
quantum state, it must have its counterpart in quantum statistical mechanics.
Since the Gamow vector belongs to an extension of the Hilbert space, its
corresponding density matix should belong to an extension of the conventional
Liouville space, called the rigged Liouville space (RLS) \ci{AGS}. Although we
can construct rigorously a dyadic product of Gamow vectors in the RLS,  these
objects do not satisfy the minimal requirements to be an state. In particular,
objects like ${\rm tr}\,(|f_0\>\<f_0|)$ or ${\rm tr}\,(|f_0\>\<f_0|H)$, where
$|f_0\>$ is the Gamow vector, are not defined. In other words, objects like
$\<f_0|f_0\>$ and $\<f_0|H|f_0\>$, representing the normalization and the mean
value of the energy respectively of a Gamow vector, cannot be defined. We have
studied the properties of Gamow dyads in RLS in
\ci{GL}.

In statistical mechanics, states are also represented by continuous 
positive and normalized functionals on an algebra of observables \ci{S}. This is
the approach we wish to analyze in this paper. We shall construct an algebra
of observables in which the Gamow ``state'' can be defined as a continuous
functional on this algebra. This functional is characterized by its decay mode
and is also positive, but cannot be normalized (as its normalization results to
be zero). Worse of all, the expectation values of the integer powers of the
Hamiltonian, $H^n$, $n=0,1,2,\dots$, vanish. As a result of this discussion we
conclude  that the Gamow functional cannot represent a
quantum state even if we admit the existence of particles with a purely
exponential decaying mode. 

This approach does not restrict its interest to
statistical mechanics but  is also  suitable for applications to the theory of
decaying nuclei
\ci{CL}. 

To better understand the notion of Gamow functional, we need to use the
notion of rigged Hilbert space (RHS). A RHS is a triplet of spaces

$$
 {\bf\Phi}\subset{\cal H}\subset{\bf\Phi}^\times
$$
where $\cal H$ is the Hilbert space of pure normalized states of a quantum
system, $\bf\Phi$ is a space of test vectors (usually a space of functions called
the space of test functions) with its own topology which is stronger (in the
sense that has more open sets, less convergent sequences and that the canonical
injection $i:{\bf\Phi}\longmapsto{\cal H}$, $i(\vf)=\vf$, is continuous).
${\bf\Phi}^\times$ is the antidual of $\bf\Phi$ or the space of all continuous
antilinear\footnote{A functional $F$ on $\bf\Phi$ is antilinear if it is a
mapping from $\bf\Phi$ into $\C$ with the following condition:
$$
F(\alpha\,\vf+\beta\,\psi)=\alpha^*\,F(\vf)+\beta^*\,F(\psi)
$$ 
where the star denotes complex conjugation.} functionals from
$\bf\Phi$ to
$\C$. It is precisely this extension ${\bf\Phi}^\times$ of the Hilbert space
which allows the existence of generalized eigenvectors of an observable
\ci{BG}.  

This paper is organized as follows: In Section 2, we define the algebra of
observables compatible with the ``free'' or unperturbed Hamiltonian $H_0$. In
Section 3, we define the notion of states as functionals over this algebra.
In Section 4, we define the algebras of observables compatible with the
total Hamiltonian $H$ and the Gamow functionals on it. We can define these
algebras in various ways and, with an appropiate definition of the algebras, the
Gamow functionals are time reversal of each other. We close the paper with a
mathematical appendix, in which we  study the mathematical tools used
in our development.

\section{The algebra $\mathcal{A}_{0}$ of observables.}

The most intuitive model that produces quantum resonances is possibly the
resonant scattering model, in which we assume the existence of a resonant
scattering process
\ci{B}, with two dynamics. The unperturbed or free dynamics is given by $H_0$
and the perturbed dynamics by $H:=H_0+V$. We assume also that the M{\o}ller wave
operators exist and that the scattering is asymptotically complete \ci{AJS}. In
this case a theorem by Gelfand \ci{GE} and Maurin \ci{M} states that there
exists a complete set of generalized eigenvectors of $H_0$ (in a suitable RHS),
$|E\>$, for all
$E$ in the continuous spectrum of $H_0$ (which we assume to be simple and equal
to
$\Bbb R^+:=[0,\infty)$): 

$$
H_0\,|E\>=E\,|E\>\,,\hskip0.6cm E\in\mathbb R^+.
$$

The vector $|E\>$ belongs to the dual space
${\bf\Phi}^\times$ of a RHS, ${\bf\Phi}\subset{\cal H}\subset{\bf\Phi}^\times$
and the completeness means that

\begin{equation}
H_{0}=\int_{0}^{\infty }dE\,E\,|E\rangle \langle E|=\int_{0}^{\infty
}dE\int_{0}^{\infty }dE^{\prime }\,\delta (E-E^{\prime })\,E\,|E\rangle
\langle E^{\prime }|  \label{1}
\end{equation}
 Therefore, the expression (\ref{1})  for $H_0$
means that
$H_0\in {\cal L}({\bf\Phi},{\bf \Phi}^\times)$, i.e., the space of continuous
linear operators from $\bf\Phi$ into ${\bf\Phi}^\times$. See also \ci{GL}. The
action of $|E\>$ on the test function $\vf\in\bf\Phi$ gives $[\vf(E)]^*$,
the complex conjugate of the value of $\vf$ at $E$. We also have that
$\<E|\vf\>=\<\vf|E\>^*$.

Equation (\ref{1}) allows us to obtain, at least formally, the following
matrix element: 

\begin{equation}
\langle E^{\prime }|H_{0}|E^{\prime \prime }\>=\int_{0}^{\infty
}dE\,E\,\langle E^{\prime }|E\rangle \langle E|E^{\prime \prime }\>
\label{2}
\end{equation}

Since $\langle E^{\prime }|E\>=\delta (E-E^{\prime })$, where the deltas are
relative to the integration from $0$ to $\infty $, (\ref{2}) is equal to $%
E^{\prime }\delta (E^{\prime }-E^{\prime \prime })$ and, therefore, it is
well defined as a distributional kernel.

\medskip \textbf{Definition}.- An operator $O$ is said to be compatible
with $H_0$ if it has the following form: 

\begin{equation}
O=\int_0^\infty dE\,O_E\,|E\rangle\langle E|+\int_0^\infty dE\int_0^\infty
dE^{\prime}\,O_{EE^{\prime}}\,|E\rangle\langle E^{\prime}|  \label{3}
\end{equation}
where $O_E$ and $O_{EE^{\prime}}$ are ordinary functions\footnote{Here we
are using the notation in \ci{LC1,LC2,LC3,LC4,CL}.} on the variables
$E$ and $E^{\prime}$ (see Appendix).  Here, the function $O_E$ is an entire 
analytic function in a class\footnote{This class is the sum ${\cal P}+{\cal
Z}$ of the space $\cal P$ of the polynomials, considered as entire analytic
functions of a complex variable, plus the space
$\cal Z$ of entire analytic functions introduced in the Appendix.} that
contains polynomials in
$E$. The function
$O_{EE'}$ should be of the form:

\b
O_{EE'}=\sum_{ij} \lambda_{ij}\,\psi_i(E)\,\phi_j(E') \la{4}
\e
where $\psi_i(E)\,\phi_j(E')\in \cal Z$, i.e., are entire analytic functions on
the variables $E$ and $E'$ (See Appendix for a definition of ${\cal Z}$. As we
see later, this is not the only possible choice for the functions $O_{EE'}$,
although it must be, in any case functions on the complex variables $E$ and
$E'$.). The sum in (\ref{4}) is finite.

It is important to remark that the set of observables compatible with $H_{0}$
is an algebra, which we  denote as ${\cal A}_0$. See Appendix  for the
definition on the algebra operations on ${\cal A}_0$.

At this point it would be convenient to justify our choice. In fact, we want 
the following properties for the set of observables ${\cal A}_0$,
compatible with $H_0$:

i.) ${\cal A}_0$ should be an algebra. This permits the use of the
traditional point of view according to which observables form a
(topological) algebra and states are continuous, positive and normalizable
functionals on this algebra \ci{S}.

ii.) The precise choice of ${\cal A}_0$ is largely a matter of 
convenience. First of all, the set of states must contain those which are
physically meaningful. All the other criteria, seem not to be very
essential from the physical point of view.

For instance: what kind of observables should we include in ${\cal A}_0$?
Should functions on $H_0$, including $H_0$ itself, be included in ${\cal
A}_0$?

Although at the first sight one is tempted to give a positive answer to
this question, we should notice that we want to discuss the nature of
Gamow objects. These Gamow objects are supposed to describe an aspect of
resonance behaviour and resonances are assumed to be
produced in resonant scattering \ci{B}. But then, our question not always
has a positive answer in scattering theory. For instance, in the algebraic
theory of scattering developed by Amrein et al.
\ci{AMM}, the algebra
${\cal A}_0$ contains only bounded operators in the bicommutant (operators
which commute with those commuting with $H_0$) of $H_0$. Since $H_0$ is not
bounded, $H_0$ is not in the
${\cal A}_0$ of \ci{AMM}.

iii.) What is really relevant here is that the algebra of observables be
spanned by the dyads of the form $|E\>\<E|$ and $|E\>\<E'|$, where $E$ and
$E'$ run out the continuous spectrum of $H_0$. To see this, at least
intuitively, let us note that for a pair of state vectors
$\psi,\vf$ and an observable $O$, we have
$$
\<\psi|O|\vf\>=\int \<\psi|E\>\<E|O|E'\>\<E|\vf\>\,dE
$$
Then, if the kernel $\<E|O|E'\>$ satisfies the van Hove
hypothesis\footnote{This hypothesis was introduced by van Hove in his
study of unstable quantum systems.}
\ci{VH,A1}, 
$$
\<E|O|E'\>=O_E\,\delta(E-E')+O_{EE'}
$$
we have that
$$
\<\psi|O|\vf\>= \int \<\psi|E\>\<E|\vf\>\,O_E\,dE +\int
\<\psi|E\>\<E'|\vf\>\,O_{EE'}\,dE\,dE'
$$
from where (\ref{4}) follows.

Then the choice of the functions $O_E$ and $O_{EE'}$ gives the observables
that we want to consider.

iv.) As we shall see in the next section, we want to include in the
formalism states which are outside the Hilbert-Schmidt space (and therefore
are not density operators on the Hilbert space) and we have to adapt the
algebra so as to include these singular objects.

v.) Since we want Gamow objects that are continuous functionals on
operator algebras (not on ${\cal A}_0$ but instead on the derived algebras
${\cal A}_\pm$ to be defined in Section 4) and since Gamow functionals are
characterized by certain complex numbers (of the kind $E_R-i\Gamma/2$,
where $E_R$ is the resonant energy and $\Gamma$ the width \ci{B}), it
seems reasonable that the functions $O_{EE'}$ be defined over a complex
domain. Analyticity of these functions over this domain will then allow to
perform all kind of operations that are customary in the study of
resonances and Gamow vectors: contour integrals, calculus of residues, etc
\ci{B,BG,CL,AP}. 

vi.) The issue whether the algebra ${\cal A}_0$ (as well as the algebras
${\cal A}_\pm$ to be defined in Section 4) has a precise physical meaning
has the same answer as a similar question that has been addressed by the
RHS. This question is the following: given a RHS $\F\subset
\H\subset\F^\times$, what is the physical meaning of the space of test
vectors $\F$? Should $\F$ be contained or even be spanned by the space of
pure states which are physically preparable? Not neccesarily, for if
$\F$ is dense in $\H$, any physically preparable state can be approached by
a vector in $\F$ as much as we want, with respect the norm of $\H$. This
norm is produced by the scalar product, what gives the transition
amplitudes. As a matter of fact, the space $\F$, is chosen for topological
convenience as well as to determine the size of the dual space
$\F^\times$, which must contain all generalized states (like plane waves).
Thus, the specific form of the algebra ${\cal A}_0$ is also determined by
mathematical convenience.

Once we have motivated the choice of ${\cal A}_0$, let us comment some of
its properties.

It is interesting to note that the operator $O$ commutes, according to the
definition of the product in the algebra given in the Appendix, with $H_0$ 
if and only if $O_{EE'}=0$. The proof of this statement is also presented
in the Appendix. 

Also in the Appendix, we shall give the topology on the algebra ${\cal A}_0$ 
that will allow to define continuous functionals on ${\cal A}_0$. We want
to add that this topology makes the following mappings continuous:

\b
O\longmapsto O_E\hskip0.5cm;\hskip0.5cm O\longmapsto O_{EE'} \la{5}
\e
for all $E,E'\in\mathbb C$, where $\C$ is the complex plane. According to a
useful notation \ci{A1}, we can represent these two functionals as $(E|$ and
$(EE'|$ respectively, so that:

\b
O_E=(E|O) \hskip0.5cm;\hskip0.5cm O_{EE'}=(EE'|O) \la{6}
\e
which yields
\b
O=\int_0^\infty (E|O)\, |E\>\<E|\,dE+\int_0^\infty dE\int_0^\infty dE'\,
(EE'|O)\,|E\>\<E'| \la{7}
\e
This notation is consistent with the following \ci{A1}:

\b
|E\>\<E|\equiv |E) \hskip0.5cm;\hskip0.5cm |E\>\<E'|\equiv |EE') \la{8}
\e
and

\b
(E|w)=\delta(E-w) \hskip0.5cm;\hskip0.5cm (EE'|ww')=\delta(E-w)\,\delta(E'-w').
\la{9}
\e

Taking into account that $\<E|E'\>=\delta(E-E')$, where the delta refers to 
integration from $0$ to $\infty$, we  also obtain   that

\b
\<E|O|E'\>=O_E\,\delta(E-E')+O_{EE'}. \la{10}
\e
It is also important to remark that only self-adjoint elements of ${\cal
A}_0$ should be considered as observables. The condition for
self-adjointness in our case is very simple. The formal adjoint of $O$ is
given by

\begin{equation}
O^{\dagger }:=\int_{0}^{\infty }dE\,O_{E}^{*}\,|E\rangle \langle
E|+\int_{0}^{\infty }dE\int_{0}^{\infty }dE^{\prime }\,O_{EE^{\prime
}}^{*}\,|E^{\prime }\>\langle E|  \label{11}
\end{equation}

It is easy to show that this definition is consistent with the formula $
(\varphi ,O\psi )=(O^{\dagger }\varphi ,\psi )$, when $\varphi ,\,\psi \in 
\mathbf{\Phi }$ and $(-,-)$ is the scalar product on the Hilbert space
$\H$ (see (\ref14)).  Here,
$\mathbf{\Phi }$ is the space of test functions introduced earlier, on which
$|E\>$ applies.

\smallskip \textbf{Definition}.- We say that $O$ is self-adjoint if
$O=O^\dagger$. An operator $O$ of the form (\ref{3}) is an observable if and
only if it is self-adjoint.

\smallskip \textbf{Proposition}.- The operator $O$ is an observable if and only
if 

\begin{equation}
O_E=O^*_E\hskip0.7cm\mathrm{and}\hskip0.7cm O_{EE^{\prime}}=O^*_{E^{\prime}E}
\label{12}
\end{equation}
where $``^*"$ means complex conjugate.

\smallskip \textbf{Proof}.- Let us assume that $O$ is an observable. Then,
it is immediate to show that 

\begin{eqnarray}
\langle E|O|E^{\prime }\> &=&O_{E}\,\delta (E-E^{\prime })+O_{EE^{\prime }} 
\nonumber \\[2ex]
\langle E|O^{\dagger }|E^{\prime }\> &=&O_{E}^{*}\,\delta (E-E^{\prime
})+O_{E^{\prime }E}^{*}  \label{13}
\end{eqnarray}
Since $O=O^{\dagger }$, (\ref{13}) implies that $O_{E}=O_{E}^{*}$ and $%
O_{EE^{\prime }}=O_{E^{\prime }E}^{*}$. Reciprocally, if these two equations
hold, then, for any pair of test vectors $\varphi $ and $\psi $, we have
that $(\varphi ,O\psi )=(O^{\dagger }\varphi ,\psi )$, as we can easily
check. Observe that $O_{EE^{\prime }}$ is complex in general.

Now, we are more interested in clarifying the formalism we use here and the role
of quantum states on it. We do this in the next section.

\section{States.}

The theorem of Gelfand and Maurin \ci{GE,M} stablishes the existence of a RHS
$\mathbf{\Phi }\subset \mathcal{H}\subset \mathbf{\Phi }^{\times } $ such that
if $\psi,\vf\in{\bf\Phi}$, we have that

\b
(\psi,\vf)=\int_0^\infty \<\psi|E\>\<E|\vf\>\,dE \la{14}
\e
where the brackets (-,-) denote scalar product on the Hilbert space
$\mathcal{H}$. If we omit the arbitrary vector  $\psi\in\bf\Phi$ in (\ref{14}),
we have that

\begin{equation}
\varphi =\int_{0}^{\infty }|E\rangle \langle E|\varphi\rangle
\,dE  \label{15}
\end{equation}
However, formula (\ref{15}) is inconsistent as far as its right hand side is a
functional on $\mathbf{\Phi }$ (and therefore a vector in $\mathbf{
\Phi }^{\times })$ and its left hand side a vector in $\mathbf{\Phi }$. As
$\mathbf{\Phi }\subset \mathbf{\Phi }^{\times }$,  $\varphi $ can
be also looked as a vector in $\mathbf{\Phi }^{\times }$. For convenience, we
introduce the identity mapping $I$ that maps a vector on $\mathbf{\Phi }$ as the
same vector as member of $\mathbf{\Phi }^{\times }.$ This identity can
be written as:

\begin{equation}
I=\int_{0}^{\infty }|E\rangle \langle E|\,dE  \label{16}
\end{equation}

At this point, we can start the discussion on states by calculating the mean
value of a pure state $\psi$, considered as a  vector with norm one on the
Hilbert space $\cal H$, on the observable $O$. This is given by

\begin{eqnarray}
\langle \psi |O|\psi \> &=&\left[ \int_{0}^{\infty }dE\,\langle \psi
|E\rangle \langle E|\right] \,\left[ \int_{0}^{\infty }dE^{\prime
}\,O_{E^{\prime }}\,|E^{\prime }\>\langle E^{\prime }|\right.  \nonumber \\%
[0.03in]
&&+\left. \int_{0}^{\infty }dE^{\prime }\int_{0}^{\infty }dE^{\prime \prime
}\,O_{E^{\prime }E^{\prime \prime }}\,|E^{\prime }\>\langle E^{\prime \prime
}|\right] \left[ \int_{0}^{\infty }dE^{\prime \prime \prime }\,|E^{\prime
\prime \prime }\>\langle E^{\prime \prime \prime }|\psi \>\right]  \nonumber
\\[0.03in]
&=&\int_{0}^{\infty }dE\,|\langle \psi |E\>|^{2}\,O_{E}+\int_{0}^{\infty
}dE\int_{0}^{\infty }dE^{\prime }\,O_{EE^{\prime }}\,\langle \psi
|E\>\langle E^{\prime }|\psi \>  \nonumber \\[0.03in]
&=&\int_{0}^{\infty }dE\,|\psi (E)|^{2}\,O_{E}+\int_{0}^{\infty
}dE\int_{0}^{\infty }dE^{\prime }\,O_{EE^{\prime }}\,\psi ^{*}(E)\,\psi
(E^{\prime })  \label{17}
\end{eqnarray}
Obviously, this comes after $\langle \varepsilon |\zeta \rangle =\delta
(\varepsilon -\zeta )$, when $\varepsilon ,\zeta =E,E^{\prime },E^{\prime
\prime },E^{\prime \prime \prime }$. We can use here the notation $\rho_E=|\psi
(E)|^{2}$ and $\rho_{EE'}= \psi ^{*}(E)\,\psi
(E^{\prime })$. Note that $\rho_E=\rho_{EE}$.

Now, let $\rho $ be a mixture of states. Then, $\rho =\sum_{i}\lambda_{i}\,|\psi
_{i}\>\langle \psi _{i}|$ with $\sum_{i}\lambda _{i}=1$, $\lambda _{i}\ge 0$ and
$\<\psi_i|\psi_j\>=\delta_{ij}$. The mean value of the observable $O$, compatible
with
$H_{0}$, in the state
$\rho $ is given by: 

\begin{eqnarray}
\mathrm{tr}\,\rho O &=&\sum_{i}\lambda _{i}\,\langle \psi _{i}|O|\psi _{i}\>
\nonumber \\
&=&\sum_{i}\lambda _{i}\,\int_{0}^{\infty }dE\,|\psi
_{i}(E)|^{2}\,O_{E}+\sum_{i}\lambda _{i}\,\int_{0}^{\infty
}dE\int_{0}^{\infty }dE^{\prime }\,O_{EE^{\prime }}\,\psi _{i}^{*}(E)\,\psi
_{i}(E^{\prime })  \nonumber \\[0.03in]
&=&\int_{0}^{\infty }dE\,\left[ \sum_{i}\lambda _{i}\,|\psi
_{i}(E)|^{2}\,\right] O_{E}  \nonumber \\[0.03in]
&+&\int_{0}^{\infty }dE\int_{0}^{\infty }dE^{\prime }\,O_{EE^{\prime
}}\,\left[ \sum_{i}\lambda _{i}\,\psi _{i}^{*}(E)\,\psi _{i}(E^{\prime
})\right]  \label{18}
\end{eqnarray}

We call $\rho_E:=\sum_i\lambda_i\,|\psi_i(E)|^2$ and $\rho^*_{EE^{\prime}}:=
\sum_i\lambda_i$ $\psi^*_i(E)\,\psi_i(E^{\prime})$. Note that $\rho_E$ is real
and $\rho_{EE^{\prime}}$ is complex in general. It is also
true that $\rho_E=\rho_{EE}$.

Now, observe that in both cases we can write the state as 

\begin{equation}
\rho =\int_{0}^{\infty }dE\,\rho _{E}\,(E|+\int_{0}^{\infty
}dE\int_{0}^{\infty }dE^{\prime }\,\rho _{EE^{\prime }}\,(EE^{\prime }|
\label{19} 
\end{equation}
so that when applied to the observable $O$ written as

\begin{equation}
O=\int_{0}^{\infty }dE\,O_{E}\,|E)+\int_{0}^{\infty }dE\int_{0}^{\infty
}dE^{\prime }\,O_{EE^{\prime }}\,|EE^{\prime })  \label{20}
\end{equation}
gives the result

\begin{equation}
(\rho |O):=\mathrm{tr}\,\rho O=\int_{0}^{\infty }dE\,\rho
_{E}\,O_{E}+\int_{0}^{\infty }dE\int_{0}^{\infty }dE^{\prime }\,\rho
_{EE^{\prime }}\,O_{EE^{\prime }}  \label{21}
\end{equation}
where we have used the relations (\ref{9}). For the two choices (\ref{17}) and
(\ref{18}), $\rho$ in (\ref{19}) defines a continuous positive and normalized
functional on ${\cal A}_0$ and therefore a state.

At this point, we observe that the algebra ${\cal A}_0$ is a direct sum of two
subalgebras, the algebra $\frak B$ spanned by
$$
\int_0^\infty dE\, O_E \,|E)\,,\hskip1cm (|E)=|E\>\<E|)
$$
and the algebra $\frak C$ spanned by 
$$
\int_0^\infty dE\int_0^\infty dE'\, O_{EE'}\,|EE')\,,\hskip1cm
(|EE')=|E\>\<E'|).
$$
Both algebras do not have common elements other than the zero (see
Appendix). Therefore ${\cal A}_0={\frak B}+\frak C$ is a direct sum. As a
consequence, every continuous linear functionals on ${\cal A}_0$ is
the sum of a continuous linear functional on $\frak B$ plus a continuous
linear functional on $\frak C$. 

The algebras $\frak B$ and $\frak C$ are respectively  isomophic to the
algebras of the functions of the form $O_E$ and $O_{EE'}$. Therefore,
${\cal A}_0$ is isomorphic to the algebra of pairs of functions
$(O_E,O_{EE'})$ with a product that can be immediately obtained from the
product on ${\cal A}_0$.

From all this, we conclude that the most general form of a state on ${\cal
A}_0$ is of the form (\ref{19}) being $\rho_E$ and $\rho_{EE'}$ continuous
linear functionals (distributions) on the spaces of functions of the form
$O_E$ and $O_{EE'}$ respectively (see Appendix).

In this formalism, we see that there are three kind of states:

\smallskip 
i.) Pure states. A state is pure if and only if there is a square
integrable function $\psi(E)$ such that $\rho_E=|\psi(E)|^2$ and
$\rho_{EE'}=\psi^*(E)\psi(E')$.

ii.) Mixtures. For mixtures $\rho_{EE}=\rho_E$.

iii.) Generalized states, which are all others.

\smallskip
{\bf Remarks.}

i.) Pure states and mixtures have the property that $\rho_E=\rho_{EE}$.
The converse is also true, if $\rho_{EE}$ is well defined and
$\rho_E=\rho_{EE}$, then (\ref{19}) represents either a pure state or a
mixture, i.e., it admits a representation as a density operator on Hilbert
space. On the other hand, generalized states cannot be represented as a
density operator on a Hilbert space. The need for generalized states have
been established by van Hove first \ci{VH} and a mathematically consistent
definition of them was given in \ci{A1}.  Our formalism is clearly
inspired in \ci{A1}, although our goals are different as we try to
understand the role of the Gamow objects on it.

ii.) There are two kinds of generalized states, those for which
$\rho(E,E)$ is well defined in a distributional sense and those for which
does not. For example, assume that
$\rho(E,E')=\delta(E-E_0)\,\delta(E'-E_0)$. In this case, obviously
$\rho(E,E)$ does not make sense. If for a given state $\rho(E,E)$
is well defined, this is a generalized state if and only if
$\rho(E)\ne \rho(E,E)$.

\smallskip
The evolution of the state $\rho$ under the free Hamiltonian $H_0$ is

\begin{eqnarray}
(\rho _{t}|O) &=&(\rho _{0}|e^{itH_{0}}\,O\,e^{-itH_{0}})  \nonumber \\ [2ex]
&=&\int_{0}^{\infty }dE\,O_{E}\,\rho _{E}+\int_{0}^{\infty
}dE\int_{0}^{\infty }dE^{\prime }\,O_{EE^{\prime }}\,e^{it(E-E^{\prime
})}\,\rho _{EE^{\prime }}  \label{22}
\end{eqnarray}

If $O_{EE^{\prime }}$ is bounded and $\rho $ is a mixture, due to the
integrability of $\sum_{i}\lambda _{i}$ $\psi ^{*}(E)\psi (E^{\prime })$,
then $O_{EE^{\prime }}\rho _{EE^{\prime }}$ is also integrable and the
second integral term in (\ref{22}) vanishes as $t\longmapsto \infty$ as the
result of the Riemann-Lebesgue lemma. After the limit process, only the
first term remains. This fact is usually called decoherence.

\section{The algebras ${\cal A}_\pm$ of observables.}

The algebras ${\cal A}_\pm$ play the same role with respect to the total
Hamiltonian $H$ as the algebra ${\cal A}_0$ with respect to $H_0$.

First of all, let ${\bf\Omega}_\pm$ be the M{\o}ller wave operators, defined
as customary as \ci{AJS}:

$$
{\bf\Omega}_+\vf=\lim_{t\mapsto+\infty}e^{itH}\,e^{-itH_0}\,\vf=\vf^+
$$
and
$$
{\bf\Omega}_-\vf=\lim_{t\mapsto-\infty}e^{itH}\,e^{-itH_0}\,\vf=\vf^-
$$
whenever these limits exist. The M{\o}ller wave operators relate state
vectors which evolve with the total Hamiltonian $H$ with state vectors
which evolve with the free Hamiltonian $H_0$ and that are asymptotically
(as $t\longmapsto\pm\infty$) identical (in our case $\vf$ evolves freely and
$\vf^\pm$ with $H$ and
$\lim_{t\mapsto\pm\infty}(e^{-itH_0}\vf-e^{-itH}\vf^\pm)=0$).

 As the
M{\o}ller wave operators are assumed to exist, let us define \ci{BG}\footnote{If
we define ${\bf\Phi}^\pm:={\bf\Omega}_\pm \bf\Phi$, we have two new triplets
$$
{\bf\Phi}^\pm\subset{\cal H}\subset ({\bf\Phi}^\pm)^\times
$$
where $\cal H$ is the absolutely continuous part of the Hilbert space with
respect to $H$ (see \ci{AJS}). The M{\o}ller operators can be extended to
bicontinuous mappings from $\bf\Phi^\times$ into $({\bf\Phi}^\pm)^\times$, so
that (\ref{23}) makes sense. This definition is made through the duality
formula:
$$
\<\vf|E\>=\<{\bf\Omega}_\pm\vf|{\bf\Omega}_\pm|E\>=\<\vf^\pm|E^\pm\>
$$
where $\vf$ is an arbitrary vector in $\F$.}

\b
|E^\pm\>={\bf\Omega}_\pm\,|E\> \la{23}
\e
The definition (\ref{23}) makes sense as proven in \ci{BG}. Now, take $O$ as in
(\ref{3}) and write

\begin{equation}
O^{\pm }=\mathbf{\Omega }_{\pm }\,O\,\mathbf{\Omega }_{\pm }^{\dagger }\la{24}
\end{equation}
Since (\ref{23}) implies\footnote{To see this, write
$\<E|\vf\>:=\<\vf|E\>^*$ with $\vf\in\F$. Then,
$$
\<E^\pm|\vf^\pm\>=\<E|\vf\>=\<E|{\bf\Omega}_\pm^\dagger
{\bf\Omega}_\pm|\vf\>=
\<E|{\bf\Omega}_\pm^\dagger|\vf^\pm\>
$$
This expression is valid for any $\vf^\pm\F^\pm$. Then,
$\<E|{\bf\Omega}_\pm^\dagger=\<E^\pm|$ follows.} that
$\<E|{\bf\Omega}_\pm^\dagger=\<E^\pm|$, the operators
$O^\pm$ can then be written as:

\begin{equation}
O^{\pm }=\int_{0}^{\infty }O_{E}\,|E^{\pm }\rangle \langle E^{\pm
}|\,dE+\int_{0}^{\infty }dE\int_{0}^{\infty }dE^{\prime }\,|E^{\pm }\rangle
\langle E^{\prime \pm }|\,O_{EE^{\prime }}  \label{25}
\end{equation}

We say that an operator\footnote{These operators are continuous linear
functionals from ${\bf\Phi}^\pm$ into $({\bf\Phi}^\pm)^\times$. Therefore, they
are a generalization of the usual notion of operator as a linear mapping on
$\cal H$.}
$O^\pm$ is compatible with
$H$ if and only if, it can be written in the form given in equation
(\ref{25}). Since 

\b
\<E^\pm|w^\pm\>=\<E|\mathbf{\Omega }_{\pm }^{\dagger }\, \mathbf{\Omega }_{\pm
}|w\>=\<E|w\>=\delta(E-w) \la{26}
\e
we obtain that the operators of the type $O^+$ and $O^-$ in (\ref{25}) form
respective algebras that we call $\cal A_+$ and $\cal A_-$ (see in the Appendix
how to define the product for two elements of ${\cal A}_0$. After (\ref{26}), it
is clear that the product in ${\cal A}_\pm$ is defined analogously). Since
the operators ${\bf\Omega}_\pm$ are unitary\footnote{We assume asymptotic
completeness \ci{AJS}. Therefore ${\bf\Omega}_\pm$ are unitary operators
between the absolutely continuous subspaces of $H_0$ and $H$.}, the algebras
$$
{\cal A}_\pm:={\bf\Omega}_\pm\,{\cal A}_0\,{\bf\Omega}_\pm^\dagger
$$
are isomorphic (algebraically and topologically) to the algebra ${\cal
A}_0$.

 States on
these algebras have the form

\b
\rho^\pm=\int_0^\infty \rho_E \,(E^\pm|\,dE+ \int_0^\infty dE \int_0^\infty dE'\,
\rho_{EE'}\,(EE'^\pm| \la{27}
\e 
where

\b
(E^\pm|O^\pm)=O_E \hskip0.5cm;\hskip0.5cm (EE'^\pm|O^\pm)=O_{EE'} \la{28}
\e

The operators (\ref{25}) can be also written as

\b
O^\pm=\int_0^\infty dE\,O_E\,|E^\pm)+ \int_0^\infty dE \int_0^\infty dE'\,
O_{EE'}\,|EE'^\pm) \la{29}
\e
so that

\b
(E^\pm|w^\pm)=\delta(E-w) \hskip0.5cm;\hskip0.5cm (EE'^\pm|ww'^\pm)=
\delta(E-w)\,\delta(E'-w'). \la{30}
\e

This means that the operational rules in ${\cal A}_\pm$ are the same than in
${\cal A}_0$. The same can be said about the topology as the components
$O_E$ and
$O_{EE'}$ of both algebras are the same. This topology is transported from
${\cal A}_0$ to ${\cal A}_\pm$ by the M{\o}ller operators. Also pure states,
mixtures and generalized states with diagonal singularity can be written as
functionals on ${\cal A}_\pm$ exactly as on ${\cal A}_0$. Time evolution of
$\rho^\pm$ with respect to $H$ is of the form

\be
(\rho^\pm_t|O^\pm)=\int_0^\infty dE \rho_E\,O_E + \int_0^\infty dE \int_0^\infty
dE'\,e^{it(E-E')}\,\rho_{EE'}\,O_{EE'} \la{31}
\ee 
Observe that the first integral in (\ref{31})  does not evolve in
time. The second part vanishes for $t\longmapsto\pm\infty$ if
$\rho_{EE'}\,O_{EE'}$ is an integrable function in the two dimensional variable
$(E,E')$.

\section{The Gamow Functionals.}

If the pair $\{H_0,H\}$ produces resonances, these are manifested as pairs of
poles of the same multiplicity in the analytic continuation of the $S$-matrix 
in the energy representation
\ci{B} or the reduced resolvent \ci{AM}. Both are complex functions of the energy
considered as a complex variable and, under very general conditions \ci{AM}, have
poles located at the same points. These poles may have arbitrary multiplicity
and appear into complex conjugate pairs of the same multiplicity, although only
simple resonance poles yield exponentially decaying Gamow vectors \ci{AGP}. Thus,
let us assume that we have a pair of resonance poles located at the points 
$z_0=E_R-i\Gamma/2$ and its complex conjugate $z_0^*$. Within the above
formalism is quite easy to define the decaying Gamow functional.

For any function $\psi\in \cal Z$, the functional $\delta_z$ maps $\psi(E)$
into its value at $z$, $\psi(z)$. If $\phi$ is another function in $\cal Z$,
the tensor product $\delta_z\otimes \delta_{z'}$ maps the function
$\psi\otimes\phi$ into
$\psi(z)\phi(z')$.

Then, we define the decaying Gamow functional as

\b
\rho_D:= \int_0^\infty dE\int_0^\infty dE'\,\delta_{z_0^*}\otimes
\delta_{z_0}\, (EE'^+| \la{32} 
\e

This is obviously an element of $\mathcal{A}_{+}^{\times }$, the dual of the
algebra ${\cal A}_+$. Note that
$(\rho _{D})_{E}=0$ and $(\rho _{D})_{EE^{\prime }}=\delta_{z_0^*}\otimes
\delta_{z_0}$. The action of $\rho _{D}$ on $O\in \mathcal{A}_{+}$ is given by 

\begin{equation}
(\rho _{D}|O)=O_{z_{0}^{*}\,z_{0}} \la{33}
\end{equation}
The functional $\rho _{G}$ has the following properties: ($\rho _{D}(0)=\rho
_{D})$

\begin{eqnarray}
&&(\rho _{D}(t)|O)  \nonumber \\[2ex]
&=&\int_{0}^{\infty }dE\int_{0}^{\infty }dE^{\prime }\,\delta
_{z_{0}^{*}\,z_{0}}\,\,O_{EE^{\prime }}\,e^{it(E-E^{\prime
})}=e^{it(z_{0}^{*}-z_{0})}\,O_{z_{0}^{*}\,z_{0}}  \nonumber \\[2ex]
&=&e^{-t\Gamma }\,O_{z_{0}^{*}\,z_{0}}=e^{-t\Gamma }\,(\rho _{D}|O) \la{34}
\end{eqnarray}
where $z_{0}=E_{R}-i\frac{\Gamma }{2}$, being $E_{R}$ the resonant energy
and $\Gamma $ the mean life. We observe that $\rho _{D}$ decays
exponentially for all values of the time. Other properties of $\rho _{D}$
are: 

\begin{equation}
(\rho _{D}|I^{+})=0 \la{35}
\end{equation}
where $I^{+}$ is given by

$$
I^+= {\bf\Omega}_+\,I\,{\bf\Omega}_+^\dagger =\int_0^\infty |E^+\>\<E^+|\,dE
$$ 
and\footnote{Observe that $I^+$ is the canonical injection from ${\bf\Omega}_+$
into ${\bf\Omega}_+^\dagger$. See footnote 3.}
$I$ is given in (\ref{16})

\begin{equation}
(\rho _{D}|H^n)=0, \hskip0.7cm n=0,1,2,\dots \la{36}
\end{equation}
where

$$
H^n=\int_0^\infty dE\,E^n \,|E^+)
$$

We can choose the functions $O_E$ and $O_{EE'}$ in such a way that the evolution
$\rho_D(t)$ for the Gamow functional is either valid for $t>0$ only or for all
values of time. In the latter case, the
evolution law is not given by a semigroup and this eliminates the problem of
fixing the time
$t=0$ as ``the instant at which  the preparation of the quasistationary state
has been completed and starts to decay'' \ci{B,BAK}. 
In the former case,  $O_{EE'}$ cannot belong to a class of entire functions on
the variables $EE'$, as we shall see later.  

In summary, the Gamow functional $\rho_D$ has the following properties:

\smallskip
1.- It is linear and continuous on the algebra ${\cal A}_+$.

2.- It is positive, i.e., $(\rho_D|(O^+)^\dagger O^+)\ge 0$.

3.- Equation (\ref{35}) shows that the functional $\rho_D$ does not admit a
normalization\footnote{Should we have $(\rho_D|I)=\alpha\ne 0$, we could still
normalize the functional as $\frac{\rho_D}{\alpha}$.}. A quantum state is
defined to be a linear functional on an algebra, containing the observables of
the system, which is continuous, positive and normalizable. As $\rho_D$ is not
normalizable, it is not a state in the ordinary sense. In addition, equation
(\ref{36}) shows that the mean value of all powers of $H$ on $\rho_D$
vanish. This is another argument to conclude that $\rho_D$ does not
represent a truly quantum state.

Along the the decaying Gamow functional there is the growing Gamow
functional $\rho_G$ which is defined on ${\cal A}_-$ as:

\b
\rho_G= \int_0^\infty dE\int_0^\infty dE' \,\delta_{z_0}\otimes \delta_{z_0^*}
\,(EE'^-| \la{37}
\e

The growing Gamow functional $\rho_G$ has the following properties:

\smallskip
1.- The mean value of $O^-$ in $\rho_G$ is given by

\b
(\rho_G|O^-)=O_{z_0z_0^*}. \la{38}
\e

2.- It grows exponentially at all times:

\b
(\rho_G(t)|O^-)=e^{t\Gamma}\,(\rho_G|O^-) \la{39}
\e
with $\rho_G=\rho_G(0)$.

3.- It is not normalizable

\b
(\rho_G|I^-)=0 \hskip0.5cm;\hskip0.5cm I^-=\int_0^\infty |E^-\>\<E^-|\,dE.
\la{40}
\e

4.- The mean value of the energy on $\rho_G$ is zero:

\b
(\rho_G|H^n)=0, \hskip0.7cm n=0,1,2,\dots \la{41} 
\e

The relation between the algebras ${\cal A}_+$ and ${\cal A}_-$ is given by the
time reversal operator $T$. In fact , we have $T|E^\pm\>=|E^\mp\>$,
$T{\bf\Phi}^\pm={\bf\Phi}^\mp$ and $T|\phi^\pm\>=|\phi^\mp\>$ \ci{GM},
so that

\b
\<E^\pm|T|\phi^\mp\>=\<E^\pm|\phi^\pm\> = \<E|\phi\>=\<E^\mp|\phi^\mp\> \la{42}
\e
where $|\phi\>:={\bf\Omega}_+^{-1}\,|\phi^+\>= {\bf\Omega}_-^{-1}\,|\phi^-\>$  
and $|E\>={\bf\Omega}_+^{-1}\,\,|E^+\>={\bf\Omega}_-^{-1}\,|E^-\>$ \ci{BG}.
Therefore, 

\b
\<E^\pm|\,T=\<E^\mp| \la{43}
\e

Therefore, if 

$$
O^\pm=\int_0^\infty dE\, O_E\,|E^\pm\>\<E^\pm| + \int_0^\infty dE \int_0^\infty
dE' \,O_{EE'}\,|E^\pm\>\<E'^\pm|
$$
we have that

\b
T\, O^\pm\,T= \int_0^\infty dE\, O_E^*\,|E^\mp\>\<E^\mp| + \int_0^\infty dE
\int_0^\infty dE' \,O_{EE'}^*\,|E^\mp\>\<E'^\mp| \la{44}
\e
(we recall that $T\alpha|\eta\>=\alpha^*\,T|\eta\>$). Thus, we obtain

\b
{\cal A}_\pm =T\,{\cal A}_\mp\,T \la{45}
\e

The relation (\ref{44}) implies a relation between $\rho_D$ and
$\rho_G$, provided that we redefine the algebras ${\cal A}_\pm$. In the new 
${\cal A}_\pm$ the functions
$O_E$ are now polynomials on the complex variable $E$. In the new algebras 
${\cal A}_\pm$  the functions
$O_{EE'}$ will be different for ${\cal A}_+$ and for ${\cal A}_-$. For ${\cal
A}_+$, $O_{EE'}$ is of the form (\ref{4}) with

\b
\psi_i(E)\in S\cap {\cal H}^2_+ \,,\hskip0.5cm \phi_j(E')\in S\cap{\cal H}^2_-
\hskip0.5cm i,j=1,2,\dots \la{46}
\e
where $S$ is the Schwartz space\footnote{Functions in $S$ are indefinitely
differentiable at all points and they and their derivatives go to zero at
$\pm\infty$ faster than the inverse of any polynomial.} and ${\cal H}^\pm$ are
the spaces of Hardy functions on the upper half plane and the lower half plane.
Hardy functions are analytic in their respective half planes and their boundary
values on the real line are square integrable functions (see Appendix). Thus,
$O_{EE'}\in S\cap{\cal H}^2_+\otimes S\cap{\cal H}^2_-$ in the algebraic sense.

 For ${\cal A}_-$, $O_{EE'}$ is of the form (\ref{4}) with

\b
\psi_i(E)\in S\cap {\cal H}^2_- \,,\hskip0.5cm \phi_j(E')\in S\cap{\cal H}^2_+
\hskip0.5cm i,j=1,2,\dots \la{47}
\e
Thus, $O_{EE'}\in
S\cap{\cal H}^2_- \otimes S\cap{\cal H}^2_+$.

Nothing in the formalism presented so far changes with this choice except the
topology of the algebras (plus the irrelevant fact that we now have two
isomorphic ${\cal A}_0$ algebras. It is not necessary to insist in this point).
However, this choice has an interesting property: the time reversal of $\rho_D$
is
$\rho_G$ and vice versa.

Before of discussing this interesting point, it is important to remark that if
$O_{EE'}\in S\cap{\cal H}^2_+\otimes S\cap{\cal H}^2_-$, then
$e^{it(E-E')}\,O_{EE'}\in S\cap{\cal H}^2_+\otimes S\cap{\cal H}^2_-$ if and
only if $t\ge0$. The proof is given in the Appendix. Analogously, if $O_{EE'}\in
S\cap{\cal H}^2_- \otimes S\cap{\cal H}^2_+$, then 
$e^{it(E-E')}\,O_{EE'}\in S\cap{\cal H}^2_- \otimes S\cap{\cal H}^2_+$ if and
only if $t\le 0$. Thus, the time evolution for $\rho_D$ makes sense for $t\ge 0$
only and time evolution for $\rho_G$ makes sense for $t\le 0$ only. Exactly as
it happens with the Gamow vectors defined in \ci{BG}.

Let us come back to the time reversal of the Gamow functionals. For $\rho_D$ the
time reversal operation is defined as:

\b
(\rho^T_D|O^-):= (\rho_D|TO^- T) \la{48}
\e
Since

\b
T\,O^-\,T= \int_0^\infty dE\,O^*_E \,|E^+\>\<E^+| + \int_0^\infty
dE\int_0^\infty dE' \,O^*_{EE'} |E^+\>\<E'^+ \la{49}|
\e
we have that

\b
(\rho_D|TO^- T)=O^*_{z_0^* z_0} \la{50}
\e
Observe that, with this new definition, 

\b
O_{EE'}=\sum_{ij}  \vf_i(E)\,\psi_j(E') \la{51}
\e
(the coefficients $\lambda_{ij}$ in (\ref{4}) can be absorbed by the
functions
$\vf_i(E)\,\psi_j(E')$) with

\b
\vf_i(E)\in {\cal H}_-^2\cap S\hskip0.5cm;\hskip0.5cm \psi_j(E')\in {\cal
H}_+^2\cap S \la{52}
\e

After the properties of Hardy functions \ci{H}, we have that

\b
\vf^*_i(E)\in {\cal H}_+^2\cap S \hskip0.5cm;\hskip0.5cm \psi_j^*(E')\in {\cal
H}_-^2\cap S \la{53}
\e
and\footnote{This property is not true in general if $\vf_i(E),\psi_j(E')\in \cal
Z$.}

\b
\vf^*_i(z^*)=\vf(z) \hskip0.5cm;\hskip0.5cm \psi^*_j(z)=\psi(z^*) \la{54}
\e
Thus,

\b
O^*_{z_0^* z_0} =\sum_{ij}\vf^*_i(z_0^*)\,\psi^*_j(z_0) =\sum_{ij}
\vf_i(z_0)\,\psi_j(z_0^*) =O_{z_0 z_0^*} \la{55}
\e
We conclude that, for arbitrary $O^-\in{\cal A}_-$, we have

\b
(\rho_D^T|O^-) = O_{z_0 z_0^*} =(\rho_G|O^-) \la{56}
\e
Thus

\b
\rho_D^T=\rho_G \la{57}
\e
Analogously,

\b
\rho_G^T=\rho_D \la{58}
\e

We observe that the decaying Gamow functional and its mirror image act on
different algebras.

It is a belief that resonances are irreversible systems and also that it exists a
microphysical arrow of time in processes like quantum decay \ci{TDL,BH,AP}. This
belief is expressed into mathematical form by choosing the test spaces
${\bf\Phi}^\pm$ for the Gamow vectors so that time evolution is defined for the
decaying Gamow vector $|f_0\>$ for $t\ge 0$ only \ci{BG}. With our second choice
for the algebras
${\cal A}_\pm$ a similar situation occurs as the evolution group splits into two
semigroups and therefore, this picture may be also valid as a mathematical
formulation of irreversibility in decaying systems.

\section*{Acknowledgements} We thank Drs. I.E. Antoniou, A. Bohm, R. de la
Madrid, L. Lara and A. Ord\'o\~nez for enlightening discussions. Partial
financial support is acknowledged to DGICYT PB98-0370, DGICYT PB98-0360 and
the Junta de Castilla y Le\'on Project PC02/99, the CONICET and the National
University of Rosario (Argentina).

\section{Appendix}

This is a mathematical appendix in which we shall construct explicitly the
algebras ${\cal A}_0$ and ${\cal A}_\pm$ and their topologies. Due to the simple
relation between these algebras, it is enough to construct ${\cal A}_0$. For
this we have two possibilities: either
the functions $O_E$ are entire analytic or are Hardy.

The former option is simpler and the construction is as follows: Let $\cal D$ be
the space of infinitely differentiable complex functions on the set of real
numbers that have compact support. The  Fourier transform of a function in
$\cal D$ is entire analytic \ci{R}. Therefore, the space of Fourier transforms
of $\cal D$, ${\cal Z}={\cal F}({\cal D})$,  is a space of entire analytic
functions. This space has its own topology \ci{R,GS} and the product of two
functions in $\cal Z$ is another function in $\cal Z$ \ci{R}. Furthermore, the
product of a polynomial $p(z)$ times $f(z)\in\cal Z$ also belongs to $\cal Z$,
i.e., $p(z)f(z)\in\cal Z$ (which can easily derived from theorems 6.30 and 6.37
in \ci{R}).

Then, $O_E$ is a sum of a function in $\cal Z$ plus a polynomial. The two
variable function, $O_{EE'}$, has the form (\ref{4}) with $\vf_i(E)$ and
$\psi_j(E')$ in
$\cal Z$. Then, $O_{EE'}\in{\cal Z}\otimes{\cal Z}$. To show that ${\cal A}_0$ is
an algebra, let us write:

$$
G:=\int_0^\infty dE\,G_E\,|E\>\<E| + \int_0^\infty dE \int_0^\infty dE'\,G_{EE'}
\,|E\>\<E'|
$$
where $G_E$ and $G_{EE'}$ are as $O_E$ and $O_{EE'}$. Then,

\be
OG=\left\{ \int_0^\infty dE\,O_E\,|E\>\<E| + \int_0^\infty dE \int_0^\infty 
dE'\,O_{EE'}\,|E\>\<E'|\right\}\nonumber\\[2ex] 
\left\{\int_0^\infty dw\,G_w\,|w\>\<w| + \int_0^\infty dw \int_0^\infty
dw'\,G_{ww'}
\,|w\>\<w'|\right\} \nonumber\\[2ex]
= \int_0^\infty dE \int_0^\infty dw\,O_E\,G_w \,|E\>\<E|w\>\<w| \nonumber\\[2ex]
+ \int_0^\infty dE \int_0^\infty dw \int_0^\infty
dw'\, O_E\,G_{ww'}\, |E\>\<E|w\>\<w'| \nonumber\\[2ex] + \int_0^\infty dE \int_0^\infty 
dE'\, \int_0^\infty dw\, O_{EE'}\,G_w\, |E\>\<E'|w\>\<w| \nonumber\\[2ex] + 
\int_0^\infty dE \int_0^\infty  dE' \int_0^\infty dw \int_0^\infty
dw'\,O_{EE'}\,G_{ww'}\, |E\>\<E'|w\>\<w'| \nonumber\\[2ex]
= \int_0^\infty dE\,O_E\,G_E\, |E\>\<E| + \int_0^\infty dE \int_0^\infty dE'
O_E\,G_{EE'} \,|E\>\<E'| \nonumber\\[2ex]
+ \int_0^\infty dE \int_0^\infty dE' \,G_{E'}\,O_{EE'}\,|E\>\<E'|
\nonumber\\[2ex] +
\int_0^\infty dE \int_0^\infty dE' \int_0^\infty dw' \, O_{EE'}\,G_{E'w'}\,
|E\>\<w'| \la{59}
\ee

Now,  $O_E G_E$ is either a polynomial on $E$ or a function in $\cal Z$. The
functions $O_E\,G_{EE'}$ and $G_{E'}\,O_{EE'}$ are of the form (\ref{4}). Let us
take the last integral in (\ref{43}) and interchange on it $E'$ and $w'$. We
have:

\b
\int_0^\infty dE \int_0^\infty dE' \,|E\>\<E'| \int_0^\infty dw'
O_{Ew'}\,G_{w'E'} \la{60}
\e
We can immediate see that the last integral in (\ref{60}) is a function of the
form (\ref{4}). This shows that ${\cal A}_0$ is an algebra. In order to define a
topology on this algebra, we first note  that ${\cal A}_0$ considered as a vector
space  is the direct sum of three spaces:

\b
{\cal P}+{\cal Z}+{\cal Z}\otimes {\cal Z} \la{61}
\e
where $\cal P$ is the space of polynomials on the complex variable $E$. Let us
topologize $\cal P$ as follows: consider the space of all functions $f(E)\in
L^2[0,\infty)$ such that

\b
\int_0^\infty|p(E)\,f(E)|^2\,dE<\infty \la{62}
\e
This space is dense in $L^2[0,\infty)$. For each function $f(E)$ of this kind,
we define on $\cal P$ the following seminorm:

\b
q_{f,K}(p):= \sqrt{\int_0^\infty|p(E)\,f(E)|^2\,dE}+ \sup_{E\in
K}|p(E)|\,,\hskip0.6cm
\forall\,p\in
\cal P \la{63}
\e
$K$ being a compact set in $\C$.

The topologies in $\cal Z$ \ci{R} and in ${\cal Z}\otimes {\cal Z}$ \ci{P} are
standard, so that for any $p(E)+O_E+O_{EE'}\in {\cal P}+{\cal Z}+{\cal Z}\otimes
{\cal Z}$, a typical seminorm $\pi$ is of the form

\b
\pi(p(E)+O_E+O_{EE'})= q_{f,K}(p)+q(O_E) +r(O_{EE'}) \la{64}
\e
where $q$ is a seminorm in $\cal Z$ and $r$ a seminorm in ${\cal Z}\otimes
{\cal Z}$. 

Observe that not all quantum pure states are now allowed but only those
satisfying (\ref{63}). This is quite natural as condition (\ref{63}) is
fulfilled by the states in the domain of $H^n$, $n=0,1,2,\dots$ only. A
similar restriction is required for mixtures.

Now, the topology on the algebras ${\cal A}_\pm$ goes exactly as for ${\cal
A}_0$, since these algebras are isomorphic by construction. 

Functionals as $(E^+|$,  $(EE'^+|$ and  $\rho_D$ are continuous in ${\cal
A}_+$ as $(E^-|$,  $(EE'^-|$ and  $\rho_G$ are continuous in ${\cal A}_-$.
The proof is technical and we omit it here.

The second possibility for  the algebras ${\cal A}_\pm$ has been already
presented (see formulas (\ref{46}) and (\ref{47})). We want to add a few
remarks. 

1.- A Hardy function $\phi(z)$ in the upper half plane
$$\C^+:=\{z=x+iy\;\;\;;\;\;y>0\}$$ 
is a complex analytic function on $\C^+$ such that

$$
\sup_{y>0}\int_{-\infty}^\infty |\phi(x+iy)|^2\,dx =K<\infty
$$
The function $\phi(z)$ has boundary values on the real axis that determine a
square integrable function $\phi(x)$ with

$$
\int_{-\infty}^\infty |\phi(x)|^2\,dx\le K
$$
A Hardy function on the upper half plane is uniquely determined by the function
of its boundary values on the real axis \ci{H,D,K}. The space of such functions
is denoted by ${\cal H}_+^2$ and we have that ${\cal H}_+^2\subset L^2(\R)$. A
similar definition goes for Hardy functions on the lower half plane. The space
of these functions is denoted as ${\cal H}_-^2$. We have that \ci{H,D,K}

$$
{\cal H}_+^2 \oplus {\cal H}_-^2=L^2(\R)
$$

\smallskip
2.- The algebra ${\cal A}_\pm$ is now isomorphic to ${\cal P}+({\cal H}_\pm^2\cap
S)\otimes({\cal H}_\mp^2\cap S)$ and its product is defined as in (\ref{59}). The
topology in ${\cal H}_\pm^2\cap S$ is the inherited from $S$ \ci{BG}.

\smallskip
3.- Let $O_{EE'}\in S\cap{\cal H}^2_+\otimes S\cap{\cal H}^2_-$. Then, 

$$
e^{it(E-E')}\,O_{EE'}=\sum_{ij} e^{itE}\vf_i(E)\,e^{-itE'}\psi_j(E')
$$
If $t>0$, $e^{itE}\vf_i(E)\in S\cap{\cal H}^2_+$, if $\vf_i(E)\in S\cap{\cal
H}^2_+$. Also, $e^{-itE'}\psi_j(E')\in S\cap{\cal H}^2_-$, if $\psi_j(E')\in
S\cap{\cal H}^2_-$ \ci{BG}. Both properties are true if and only if $t\ge
0$ \ci{BG}. 

\medskip
Finally, let us prove that $O$ ($O^\pm$) commutes with $H_0$ ($H$), if and only
if $O_{EE'}=0$.

\begin{eqnarray}
H_{0}O &=&\left[ \int_{0}^{\infty }dE\,E\,|E\rangle \langle E|\right]
\,\left[ \int_{0}^{\infty }dE^{\prime }\,O_{E^{\prime }}\,|E^{\prime
}\>\langle E^{\prime }|\right.  \nonumber \\[2ex]
&+&\left. \int_{0}^{\infty }dE^{\prime }\int_{0}^{\infty }dE^{\prime \prime
}\,O_{E^{\prime }E^{\prime \prime }}|E^{\prime }\>\langle E^{\prime \prime
}|\right]  \nonumber \\[2ex]
&=&\int_{0}^{\infty }dE\int_{0}^{\infty }dE^{\prime }\,E\,O_{E^{\prime
}}\,|E\rangle \langle E|E^{\prime }\>\langle E^{\prime }|  \nonumber \\[2ex]
&+&\int_{0}^{\infty }dE\int_{0}^{\infty }dE^{\prime }\int_{0}^{\infty
}dE^{\prime \prime }\,E\,O_{E^{\prime }E^{\prime \prime }}\,|E\rangle
\langle E|E^{\prime }\>\langle E^{\prime \prime }|  \la{65}
\end{eqnarray}

Since $\<E|E^{\prime }\>=\delta (E-E^{\prime })$, (\ref{65}) finally gives 

\be
H_0 O=\int_{0}^{\infty }dE\,E\,O_{E}\,|E\rangle \langle E|+\int_{0}^{\infty
}dE\int_{0}^{\infty }dE^{\prime \prime }\,E\,O_{EE^{\prime \prime
}}\,|E\>\langle E^{\prime \prime }|  \nonumber
\ee

We analogously prove that 

\begin{eqnarray}
OH_{0} &=&\int_{0}^{\infty }dE\,E\,O_{E}\,|E\rangle \langle
E|+\int_{0}^{\infty }dE\int_{0}^{\infty }dE^{\prime \prime
}\,E\,O_{E^{\prime \prime }E}\,|E^{\prime \prime }\>\langle E|  \nonumber \\
&=&\int_{0}^{\infty }dE\,E\,O_{E}\,|E\rangle \langle E|+\int_{0}^{\infty
}dE^{\prime \prime }\int_{0}^{\infty }dE\,E^{\prime \prime }\,O_{EE^{\prime
\prime }}\,|E\>\langle E^{\prime \prime }|  \label{6}
\end{eqnarray}

Therefore, $H_{0}O=OH_{0}$ if and only if $EO_{EE^{\prime \prime
}}=E^{\prime \prime }O_{EE^{\prime \prime }}$ This implies that $
(E-E^{\prime \prime })O_{EE^{\prime \prime }}=0$ and since $O_{EE^{\prime
\prime }}$ is nonsingular, we conclude that $O_{EE^{\prime \prime }}=0$.
Reciprocally, if $O_{EE^{\prime \prime }}=0,$ then, $H_{0}$ and $O$ commute
Therefore, an operator $O$ commutes with $ H_{0} $ if. and only if
$O_{EE^{\prime \prime }}=0$. As in the general case, 
$O_{EE^{\prime \prime }}\neq 0$, we conclude that $\mathcal{A}_{0}$ is a
noncommutative algebra.The same result is obtained if we replace $H_0$ by $H$
and ${\cal A}_0$ by ${\cal A}_\pm$.

\end{document}